\documentclass[twocolumn,showpacs,floatfix,aps]{revtex4}

\usepackage{graphicx}
\usepackage{dcolumn}
\usepackage{bm}
\usepackage[T1]{fontenc}     

\begin{document}

\title{Recent searches for solar axions and large extra dimensions}

\author{R. Horvat} 
\author{M. Kr\v{c}mar} 
\author{B. Laki\'{c}} 
\affiliation{Ru\dj er Bo\v{s}kovi\'{c} Institute, P.O.Box 180, 10002 Zagreb,
             Croatia}

\date{\today}

\begin{abstract}
We analyze the data  from two recent experiments designed to search for solar
axions within the context of multidimensional theories of the Kaluza-Klein 
type. In these experiments, axions were supposed to be emitted from the solar
core, in $M1$ transitions between the first excited state and the
ground state of $^{57}$Fe and $^{7}$Li. Because of the high multiplicity of
axionic Kaluza-Klein states which couple with the strength of ordinary 
QCD axions,
we obtain much more stringent experimental limits on the four-dimensional
Peccei-Quinn breaking scale $f_{\rm PQ}$, compared with the solar QCD axion
limit. Specifically, for the $^{57}$Fe experiment, $f_{\rm PQ} \agt 1 \times 
10^{6}\;{\rm GeV}$ in theories with two extra 
dimensions and a higher-dimensional gravitational scale $M_H$ of order 
100 TeV, and $f_{\rm PQ} \agt 1 
\times 10^{6}\;{\rm GeV}$ in theories with three extra 
dimensions and $M_H $ of order 1 TeV (to be compared with 
the QCD axion limit, $f_{\rm PQ} \agt 8 \times 10^{3}\;{\rm GeV}$). 
For the $^{7}$Li 
experiment, $f_{\rm PQ} \agt 1.4 \times 10^{5}\;{\rm GeV}$ and $ 3.4 \times
10^{5}\;{\rm GeV}$, respectively (to be compared with the  
QCD axion limit, $f_{\rm PQ} \agt 1.9 \times  10^{2}\;{\rm GeV}$). It is an 
interesting feature of our results that, in most cases, the
obtained limit on $f_{\rm PQ}$ cannot be coupled with the mass of the axion,
which is essentially set by the (common) radius of the extra dimensions. 
\end{abstract}


\maketitle

\textit{Introduction.}---The recognition that there may be one or 
more extra space dimensions which
have sizes of the order of a millimeter raises a good deal of interest 
\cite{Ark98}. The main motivation for introducing new physics is the 
possibility to associate the Fermi scale with the Planck scale of 
(4 + $n$)-dimensional
gravity, thereby providing a natural solution for the hierarchy problem of
particle physics. Specifically, it has been suggested in such a novel
theoretical framework that ordinary matter is confined to a 3-brane
configuration, while gravity and other, hypothetical, particles which
interact very weakly with matter (such as sterile neutrinos or axions)
are only allowed to inhabit the extra compactified dimensions. 

The fact that the current validity of Newton's law of gravitation ($\sim
0.1\;{\rm mm}$) allows for the existence of extra dimensions
as large as millimeters made these models attractive for phenomenology
\cite{Hoy00}. Moreover, because of the presence of a large number of new
particles at the TeV scale, one can expect a rich phenomenology
for collider physics. Finally, if future collider experiments fail to
discover supersymmetry, which is the almost universally accepted framework for
constructing extensions of the standard model and the main proposal that
solves the hierarchy problem, the whole scenario might become an exceedingly
interesting alternative.

In spite of all the attractive features mentioned above, it is clear that
the simplest models with TeV fundamental scales and large extra
dimensions will not work any more in all particle physics phenomena that
invoke high-energy scales. Thus, for instance, mechanisms for neutrino mass
generation (the seesaw mechanism), gauge coupling unification, proton decay,
and the Peccei-Quinn (PQ) mechanism for explanation of the strong 
$CP$ problem in the quantum chromodynamics (QCD) need to be
revised in the new theory. In the latter case, in order to preserve the
axion solution to the strong $CP$ problem, three independent
higher-dimensional features leading to an ``invisible'' axion were
discovered \cite{Die00}.

In this paper, in light of this novel framework,we review data from two
recent experiments \cite{Krc98,Krc01} aimed to detect solar, 
near-monochromatic hadronic axions. 
The experimental efforts \cite{Krc98,Krc01} to detect solar 
axions were the 
first to employ a new scheme for detection of hadronic  
axions which couple only to nucleons, as
proposed by Moriyama \cite{Mor95} and by Kr\v{c}mar and collaborators 
\cite{Krc01}. Having been convinced of a mechanism for
achieving an invisible axion \cite{Die00}, one is then allowed to assume an
infinite tower of Kaluza-Klein (KK) excitations as the most important 
feature of
placing the QCD axion in the extra dimensions. Our kinematical limits are
high enough to ensure  high multiplicity of the KK axion modes 
(for two or three
extra dimensions), leading to a much larger flux than in ordinary theories and
enabling us to place a much more stringent limit on the PQ-breaking scale
$f_{\rm PQ}$. In addition, our results on $f_{\rm PQ}$ 
practically demonstrate the
most important  feature leading to the ``invisibility'' of the axion: a
complete decoupling of the mass of the axion from the PQ-breaking scale
$f_{\rm PQ}$.

\textit{QCD axions.}---Peccei-Quinn solution to the strong $CP$ 
problem in QCD \cite{Pec77}
predicts the existence of a neutral, spin-zero pseudoscalar particle, called
the (QCD) axion, associated with the spontaneously broken PQ symmetry. 
The mass of the axion ($m_{\rm PQ}$) is related to the 
PQ symmetry breaking scale
by $m_{\rm PQ}/{\rm eV} = 6 \times 10^6~{\rm GeV}/f_{\rm PQ}$. 
Two classes of axion models are discussed commonly in the literature:    
KSVZ (Kim-Shifman-Vainshtein-Zakharov) models \cite{Kim79}
and DFSZ (Dine-Fischler-Srednicki-Zhitnitski\u{i}) or grand
unified theory (GUT) models \cite{Din81}. Since the KSVZ axions do not 
interact with electrons at tree level, they are referred to as hadronic 
axions. In addition, their coupling to photons, 
$g_{a\gamma\gamma}\!\propto\!m_{\rm PQ}\,[\,E/N\!-\!2(4\!+\!z\!+\!w)/
3(1\!+\!z\!+\!w)\,]$, depends upon a model-dependent numerical parameter 
$E/N$, while for DFSZ axions $E/N = 8/3$. As such, hadronic models with 
$E/N=2$ \cite{Kap85} will have greatly suppressed photon couplings because of
a cancellation of two unrelated numbers, one being a function of the number 
of new quarks and their charges, the other a function of quark mass
ratios $z\! \equiv\! m_u/m_d\! \simeq\! 0.55$ and
$w\! \equiv\! m_u/m_s\! \simeq\! 0.029$
\cite{Leu96}.
 
Cosmological and astrophysical considerations appear to
restrict the PQ-breaking scale to two possible ranges, the so-called 
``hadronic axion window'' of hot dark matter interest \cite{Moro98}, 
$3 \times 10^5\;{\rm GeV} 
\alt f_{\rm PQ} \alt 7 \times 10^5\;{\rm GeV}$ \cite{Raf96,Eng90}, 
and a window of cold dark matter
interest, $10^9\;{\rm GeV}\alt f_{\rm PQ} \alt 10^{12}\;{\rm GeV}$ 
\cite{Raf99}. 
The first range exists only for hadronic QCD axions as long as the axion to
photons coupling is sufficiently small ($E/N = 2$). DFSZ-type QCD axions are
excluded from this region by the globular-cluster arguments based on 
the axion to photons and
axion to electron couplings \cite{Raf96}. These arguments do not affect 
hadronic QCD axions which
couple only to nucleons ($E/N = 2$) because their interactions with photons 
and electrons are strongly suppressed.

It should be noted that the hadronic axion window is indicated by the
supernova (SN) 1987A cooling and axion burst arguments which suffer from
statistical weakness, with only 19 neutrinos being observed, as well as 
from all uncertainties related to the axion emission from a hot and dense
nuclear medium \cite{Raf96,Eng90}. 
It is therefore of crucial importance to experimentally
measure or constrain the hadronic axion window.

In the following, the questions concerning the PQ-breaking scale related
to the hadronic axions will be addressed.

\textit{Limits on $f_{\rm PQ}$ from $^{57}${\rm Fe} and $^7${\rm Li} 
experiments.}---Two
new sources of near-monochromatic axions which might be emitted from the 
solar core have been recently proposed: (i) thermally excited nuclei of
$^{57}$Fe which is one of the stable isotopes of iron,  
exceptionally abundant among the heavy elements in the Sun \cite{Mor95}, 
and (ii) excited nuclei of $^7$Li
produced in the solar interior by $^7$Be electron capture and
thus accompanying the emission of $^7$Be solar neutrinos of energy 
384 keV \cite{Krc01}.
Since the axion is a pseudoscalar particle, one can expect the emission of
near-monochromatic axions during $M1$ transitions
between the first excited level and the ground state in $^{57}$Fe and $^7$Li.
The high temperatures in the center of the Sun ($\sim\!1.3$ keV) 
symmetrically
broaden the axion line to a full width at half maximum
of about 5 eV and 0.5
keV owing to the motion of axion emitters $^{57}$Fe and $^7$Li, 
respectively. As a result of Doppler broadening, these   
axions, approximately centered at the
transition energy $E_{0}\!=\!14.4$ keV and $E_{0}\!=\!478$ keV, would be
resonantly absorbed by the same nucleus $^{57}$Fe and $^7$Li, respectively,
in a laboratory on the Earth. The detection of subsequent emission of
gamma rays either of 14.4 keV or 478 keV would be a sign of axion existence. 

Following the calculations in Refs.\ \cite{Krc01,Mor95}, one can find that
the rate of excitation
per particular nucleus which is expected for solar-produced axions 
incident on a laboratory target of that metal is given by 
 \begin{equation}
      P(m_{\rm PQ}) = \int_{-\infty}^{+\infty}dE_a\,\frac{d\Phi(E_a)}{dE_a}\,
            \sigma_D(E_a)\;,
   \label{eq1}
\end{equation}
where $d\Phi(E_a)/dE_a$ is the differential flux of particular solar axions 
and $\sigma_D(E_a)$ is the effective cross section for resonant absorption
of these axions by a particular nucleus on the Earth. 
Both 
$d\Phi(E_a)/dE_a$ and $\sigma_D(E_a)$ increase as 
$\Gamma_a/\Gamma_{\gamma}$. Here 
\begin{equation}
  \frac{\Gamma_a}{\Gamma_{\gamma}} = \left( \frac{k_a}{k_{\gamma}}
  \right)^{3} \frac{1}{2\pi\alpha}\; \frac{1}{1 + \zeta^{2}}
  \left[ \frac{g_{0}\beta + g_{3}}
  {\left( \mu_{0} - \frac{1}{2} \right)\beta + \mu_{3} - \eta} \right]^{2}
  \label{eq2}
\end{equation}
represents the branching ratio of the $M$1 axionic transition
relative to the gamma transition \cite{Avi88} and contains
the nuclear-structure-dependent terms
$\beta$ and $\eta$ as well as the isoscalar and isovector nuclear 
magnetic moments $\mu_{0}$ and $\mu_{3}$. 
The momenta of the photon and the axion are denoted by $ k_{\gamma} \approx 
E_0 $ and $ k_a \approx \sqrt{E_{0}^{2} - m_{\rm PQ}^{2}} $, 
respectively, while 
$\alpha = 1/137$ is the fine structure
constant, and $\zeta$ is the $E$2/$M$1 mixing ratio.
The isoscalar and isovector axion-nucleon coupling constants, 
$g_{0}$ and $g_{3}$,
are related to $f_{\rm PQ}$ in the hadronic axion model \cite{Kap85,Sre85} by
the expressions
 \begin{equation}
      g_{0} = -\frac{m_N}{f_{\rm PQ}}\,
                 \frac{1}{6}\,
                 \left[\,2S+\left(3F-D\right)\frac{1+z-2w}{1+z+w}\,\right]
      \label{eq3} 
 \end{equation}
and
 \begin{equation}
      g_{3} = -\frac{m_N}{f_{\rm PQ}}\,\frac{1}{2}\,
                \left(D+F\right)\frac{1-z}{1+z+w}\;,
      \label{eq4}
 \end{equation}
where $m_N$ is the nucleon mass, the constants $F$ and
$D$ are the invariant matrix elements of the axial current, determined from 
hyperon semi-leptonic decays, and $S$ is the flavor-singlet axial-vector
matrix element extracted from polarized structure function data. 

The experimental methods as described above are based on 
axion to nucleon coupling,
both at the source as well as at the detector, and therefore favorable for
investigating the hadronic axions. First experiments performed along this
new line of solar axion searches set an upper
limit on hadronic axion mass of 745 eV ($^{57}$Fe experiment) \cite{Krc98} and
of 32 keV ($^7$Li experiment) \cite{Krc01} at the 95\% confidence level. 
Translating these results into limits on the PQ-breaking scale, one obtains
$f_{\rm PQ} \agt 8 \times 10^{3}\;{\rm GeV}$ and $f_{\rm PQ} \agt 1.9 \times 
10^{2}\;{\rm GeV}$ for the experiment with $^{57}$Fe and $^7$Li, respectively.

\textit{Axions in large extra dimensions.}---In Ref.~\cite{Ark98}, 
it was noted that 
with $n $ compact extra dimensions, and factorizable geometry with volume 
$V_n $, the relation between the familiar Planck scale $M_{\rm Pl} = 1.22
\times 10^{19}~{\rm GeV}$ and the
higher-dimensional gravitational scale $M_H $ is given by the formula
\begin{equation}
      M_{\rm Pl}^{2} = M_{H}^{n+2} \, V_{n}\;,
      \label{eq5}
\end{equation}
where $V_n \equiv R^n $ was considered to be exponentially large, such that
$M_H \sim M_W $. Similarly, one can place the QCD axion in the ``bulk'' of
$\delta $ extra dimensions \cite{Ark98,Die00,Cha99}, by considering a relation 
of the type Eq.~(\ref{eq5}),
\begin{equation}
      f_{\rm PQ}^2 = \bar{f}_{\rm PQ}^2 \, M_S^{\delta} \, V_{\delta}\;,
 \label{eq6}
\end{equation}
now connecting the four-dimensional PQ-breaking scale $f_{\rm PQ} $ with a
higher-dimensional PQ-breaking scale $\bar{f}_{\rm PQ}$, and $M_S $ is the
string scale, $M_S \sim M_H $. 
The most restrictive limits on the compactification scale $M_H$ for $n$=2 and 3
extra dimensions come from astrophysics (for the most stringent constraints,
see the recent work \cite{Han01}).

Since the astrophysical
limits on $f_{\rm PQ} $ \cite{Raf96,Eng90,Raf99} are many orders of 
magnitude larger than $M_H \sim M_W $,  
one should account for such a large mass scale by
introducing the axion field in higher dimensions, with $\bar{f}_{\rm PQ}$ that
could even be much smaller than $\sim $ $M_W $ \cite{Dil00}. On the other 
hand, as $f_{\rm PQ} << M_{\rm Pl}$, it is natural to 
assume that $\delta \leq n$
\cite{Die00,Dil00}. The full generalization of the PQ mechanism to higher
dimensions can be found in Ref.~\cite{Die00}.

Another feature of the higher-dimensional axion field important to us 
is its Kaluza-Klein decomposition. These four-dimensional modes with an
almost equidistant mass-splitting of order $1/R $ will be emitted from
excited nuclei of $^{57}$Fe and $^{7}$Li up to their kinematic limits. Only
the zero mode transforms under the PQ transformation as the true
axion, and therefore is only required to have a derivative coupling to
fermions, thereby playing a role of the ordinary QCD axion. It is, however,
to the  higher-dimensional structure of the axion field that each KK mode
has identical derivative couplings to fermions, 
with strength set by $f_{\rm PQ}
$ (the coupling strength of the full linear superposition of the KK states
is set by $\bar{f}_{\rm PQ}$).   

Now, we calculate the rate from Eq.~(\ref{eq1}) as a function of 
the KK axion mass, with
$k_a \approx \sqrt{E_{0}^2 - m_{\vec{\delta}}^2 }$. The masses of the 
KK modes are given by
\begin{equation}
      m_{\vec{\delta}} = \frac{1}{R} \sqrt{ n_1^2 + n_2^2 + ...  
                        + n_{\delta}^2}
                       \equiv \frac{|\vec{\delta}|}{R}\;,
    \label{eq7}
\end{equation}
where we assume that all $n$ extra dimensions are 
of the same size $R$. As a
next step we need to calculate a sum due to contributions of the massive
KK modes. Because of the smallness of the mass splitting for the size $R$
large enough $(\sim 1/R)$, it is justifiable to use integration instead of
summation \cite{Gui99}. The rate of excitation per particular nucleus 
from Eq.~(\ref{eq1}) therefore reads 
\begin{equation}
      P = \frac{ 2 \, {\pi}^{\delta / 2}}{\Gamma (\delta / 2)} \,
          R^{\delta} \, \int_{0}^{E_0}dm \, m^{\delta -1} P(m) \; .
    \label{eq8}
\end{equation}
For $n=2,3,4 $ extra dimensions $(\delta \leq n )$, our limits on $f_{\rm PQ}$
for both experiments are summarized in 
Tables~\ref{tab:table1} and \ref{tab:table2}.

\begin{table}[!tbp] 
\caption{\label{tab:table1} Limits on $f_{\rm PQ}$, 
$m_{\rm PQ}$, and $m_a$ derived 
from the experiment with $^{57}$Fe \cite{Krc98} when the QCD axion is placed 
in the bulk of two and three extra spacetime dimensions.}
\begin{ruledtabular}
\begin{tabular}{ccccccc}
 $n=2$
 &\multicolumn{3}{c}{$M_H = 100$ TeV}&\multicolumn{3}{c}{$M_H = 1000$ TeV}\\
 $R$
 &\multicolumn{3}{c}{$1.22\times 10^3~{\rm keV}^{-1}$}
 &\multicolumn{3}{c}{$12.2~{\rm keV}^{-1}$}\\
 $1/2 \, R^{-1}$
 &\multicolumn{3}{c}{0.4~eV} &\multicolumn{3}{c}{41~eV}\\
 $\delta$
         &\multicolumn{2}{c}{1}
         &2
         &\multicolumn{2}{c}{1}
         &2\\
 \hline
 $f_{\rm PQ}/{\rm GeV}\agt$
 &\multicolumn{2}{c}{$9\times 10^4$} 
 &$1\times 10^6$ 
 &\multicolumn{2}{c}{$3\times 10^4$} 
 &$1\times 10^5$\\
 $m_{\rm PQ}/{\rm eV}\alt$
 &\multicolumn{2}{c}{67}
 &6
 &\multicolumn{2}{c}{211}
 &60\\
 $m_a/{\rm eV}\alt$
 &\multicolumn{2}{c}{0.4}
 &0.4
 &\multicolumn{2}{c}{41}
 &41\\
 \hline
  \hline
 $n=3$
 &\multicolumn{3}{c}{$M_H = 1$ TeV}&\multicolumn{3}{c}{$M_H = 10$ TeV}\\
 $R$
 &\multicolumn{3}{c}{$53~{\rm keV}^{-1}$}
 &\multicolumn{3}{c}{$1.14~{\rm keV}^{-1}$}\\
 $1/2 \, R^{-1}$
 &\multicolumn{3}{c}{9.4~eV} &\multicolumn{3}{c}{439~eV}\\
 $\delta$&1&2&3&1&2&3\\
 \hline
 $f_{\rm PQ}/{\rm GeV}\agt$
 &$4\!\times \!10^4$ &$2\!\times \!10^5$ &$1\!\times \!10^6$
 &$2\!\times \!10^4$ &$3\!\times \!10^4$ &$6\!\times \!10^4$\\
 $m_{\rm PQ}/{\rm eV}\alt$
 &146&29&6&375&197&103\\
 $m_a/{\rm eV}\alt$
 &9.4&9.4&6&375&197&103
\end{tabular}
\end{ruledtabular}
\end{table}
\begin{table}[!tbp]
\caption{\label{tab:table2} Limits on $f_{\rm PQ}$, $m_{\rm PQ}$, 
and $m_a$ derived 
from the experiment with $^7$Li \cite{Krc01} when the QCD axion is placed in 
the bulk of two, three and four extra spacetime dimensions.}

\begin{ruledtabular}
\begin{tabular}{ccccccc}
 $n=2$
 &\multicolumn{3}{c}{$M_H = 100$ TeV}&\multicolumn{3}{c}{$M_H = 1000$ TeV}\\
 $R$
 &\multicolumn{3}{c}{$1.22\times 10^3~{\rm keV}^{-1}$}
 &\multicolumn{3}{c}{$12.2~{\rm keV}^{-1}$}\\
 $1/2 \, R^{-1}$
 &\multicolumn{3}{c}{0.4~eV} &\multicolumn{3}{c}{41~eV}\\
 $\delta$
 &\multicolumn{2}{c}{1}
 &2
 &\multicolumn{2}{c}{1}
 &2\\
 \hline
 $f_{\rm PQ}/{\rm GeV}\agt$
 &\multicolumn{2}{c}{$5\times 10^3$} 
 &$1\times 10^5$ 
 &\multicolumn{2}{c}{$2\times 10^3$} 
 &$1\times 10^4$\\
 $m_{\rm PQ}/{\rm eV}\alt$
 &\multicolumn{2}{c}{$1\times 10^3$}
 &43
 &\multicolumn{2}{c}{$4\times 10^3$}
 &429\\
 $m_a/{\rm eV}\alt$
 &\multicolumn{2}{c}{0.4}
 &0.4
 &\multicolumn{2}{c}{41}
 &41\\
 \hline
  \hline
 $n=3$
 &\multicolumn{3}{c}{$M_H = 1$ TeV}&\multicolumn{3}{c}{$M_H = 10$ TeV}\\
 $R$
 &\multicolumn{3}{c}{$53~{\rm keV}^{-1}$}
 &\multicolumn{3}{c}{$1.14~{\rm keV}^{-1}$}\\
 $1/2 \, R^{-1}$
 &\multicolumn{3}{c}{9.4~eV} &\multicolumn{3}{c}{439~eV}\\
 $\delta$&1&2&3&1&2&3\\
 \hline
 $f_{\rm PQ}/{\rm GeV}\agt$
 &$2\!\times \!10^3$ &$3\!\times \!10^4$ &$3\!\times \!10^5$
 &$9\!\times \!10^2$ &$4\!\times \!10^3$ &$2\!\times \!10^4$\\
 $m_{\rm PQ}/{\rm eV}\alt$
 &$3\!\times \!10^3$&211&18&$7\!\times \!10^3$&$1\!\times \!10^3$&316\\
 $m_a/{\rm eV}\alt$
 &9.4&9.4&9.4&439&439&316\\
 \hline
  \hline
 $n=4$
 &\multicolumn{6}{c}{$M_H = 1$ TeV}\\
 $R$
 &\multicolumn{6}{c}{$0.11~{\rm keV}^{-1}$}\\
 $1/2 \, R^{-1}$
 &\multicolumn{6}{c}{$4.5\times 10^3~{\rm eV}$}\\
 $\delta$
 &\multicolumn{2}{c}{1}
 &2
 &\multicolumn{2}{c}{3}
 &4\\
 \hline
 $f_{\rm PQ}/{\rm GeV}\agt$
 &\multicolumn{2}{c}{$5\times 10^2$} 
 &$1\times 10^3$ 
 &\multicolumn{2}{c}{$3\times 10^3$} 
 &$8\times 10^3$\\
 $m_{\rm PQ}/{\rm eV}\alt$
 &\multicolumn{2}{c}{$1\times 10^4$} 
 &$5\times 10^3$ 
 &\multicolumn{2}{c}{$2\times 10^3$} 
 &714\\
 $m_a/{\rm eV}\alt$
 &\multicolumn{2}{c}{$4\times 10^3$} 
 &$4\times 10^3$ 
 &\multicolumn{2}{c}{$2\times 10^3$} 
 &714
\end{tabular}
\end{ruledtabular}
\end{table}

\textit{Discussion.}---One notices from
Tables~\ref{tab:table1} and \ref{tab:table2} that our lower limits on
$f_{\rm PQ}$ are always much more stringent than that obtained in conventional
theories. Obviously, this is due to the fact that for $E_0 > R^{-1}$ the
multiplicity of states which can be produced is large. Going to higher $n$
the mass splitting of the spectrum becomes larger, thereby decreasing the
multiplicity, and the bound on $f_{\rm PQ}$ is less stringent. Such a behavior
is clearly displayed in Tables~\ref{tab:table1} and \ref{tab:table2}.  

Another feature visible in our 
Tables~\ref{tab:table1} and \ref{tab:table2} represents a practical
demonstration of the effect found in Ref.~\cite{Die00} that 
the mass of the axion 
can become independent of the energy scale associated with the breaking of the
PQ symmetry. Such an effect can be used to decouple the mass of the axion
from its couplings to ordinary matter, thereby providing a sought-for 
method of
rendering the axion invisible in higher-dimensional scenarios. The effect of
KK modes on the axion mass matrix is such that the zero-mode axion mass is
strictly bounded by the radius of the extra dimensions, $m_a \leq
(1/2)R^{-1}$. Thus, in higher dimensions the mass of the axion is
approximatively given as \cite{Die00} 
\begin{equation}
  m_a \approx {\rm min} \, \left( \frac{1}{2} \, R^{-1}, m_{\rm PQ} \right)\;.
    \label{eq9} 
\end{equation}
We see that for most combinations of $n, \delta $, and $R$, the upper limit
on $m_{\rm PQ}$ is considerably higher 
than $(1/2)R^{-1}$, and therefore cannot
be considered as a genuine limit on the mass of the axion.

\textit{Summary}.---We have interpreted data from two recent experiments aimed
to search for solar, near-monochromatic axions, 
assuming KK axions coming from the Sun. Within
the context of  conventional hadronic models with $E/N = 2$, both
experiments set a stringent upper limit on the axion mass. We have shown that
data, when interpreted in the higher-dimensional framework, cannot be used,
in most cases, to set any relevant limit on the axion mass. On the other
hand, our lower limits on the four-dimensional PQ-breaking scale turned out
to always be a few orders of magnitude more stringent than the ordinary
QCD limit. Finally, we stress that the most restrictive bounds we have derived
from the $^{57}$Fe experiment ($f_{\rm PQ} \agt 1 \times 10^{6}\;{\rm GeV}$ 
in theories with two extra dimensions and $M_H \sim 100~{\rm TeV}$ 
as well as in theories with three extra dimensions and $M_H \sim 1~{\rm TeV}$)
and from the $^7$Li experiment ($f_{\rm PQ} \agt 1.4 \times 10^{5}\;{\rm GeV}$ 
and $3.4 \times 10^{5}\;{\rm GeV}$, respectively) fall into the parameter
space of hot dark matter interest.


\begin{thebibliography}{100}
\bibitem{Ark98} 
     N. Arkani-Hamed, S. Dimopoulos, and G. Dvali, Phys. Lett. B {\bf 429},
     263 (1998); Phys. Rev. D {\bf 59}, 086004 (1999); I. Antoniadis, 
     N. Arkani-Hamed,
     S. Dimopoulos, and G. Dvali, Nucl. Phys. {\bf B516}, 70 (1998).
\bibitem{Hoy00} 
     C. D. Hoyle {\it et al}., Phys. Rev. Lett. {\bf 86}, 1418 (2001). 
\bibitem{Die00}  
     K. R. Dienes, E. Dudas, and  T. Gherghetta, Phys. Rev. D {\bf 62}, 
     105023 (2000).
\bibitem{Krc98}  
     M. Kr\v{c}mar {\it et al}., Phys. Lett. B {\bf442}, 38 (1998).
\bibitem{Krc01}  
     M. Kr\v{c}mar {\it et al}., Phys. Rev. D {\bf 64}, 115016 (2001).
\bibitem{Mor95}  
     S. Moriyama, Phys. Rev. Lett. {\bf 75}, 3222 (1995).
\bibitem{Pec77}
     R. D. Peccei and H. R. Quinn, Phys. Rev. Lett. {\bf
     38}, 1440 (1977); Phys. Rev. D {\bf 16}, 1791 (1977).
\bibitem{Kim79}
     J. E. Kim, Phys. Rev. Lett. {\bf 43}, 103 (1979); M. A. Shifman,
     A. I. Vainshtein, and V. I. Zakharov, Nucl. Phys. {\bf B166},
     493 (1980).
\bibitem{Din81}
     M. Dine, W. Fischler, and M. Srednicki, Phys. Lett.
     {\bf 104B}, 199 (1981); A. R. Zhitnitski\u{i}, Yad. Fiz.
     {\bf 31}, 497 (1980) [Sov. J. Nucl. Phys. {\bf 31}, 260 (1980)].
\bibitem{Kap85}
     D. B. Kaplan, Nucl. Phys. {\bf B260}, 215 (1985).
\bibitem{Leu96}
     H. Leutwyler, Phys. Lett. B {\bf 378}, 313 (1996).
\bibitem{Moro98}
     T. Moroi and H. Murayama, Phys. Lett. B {\bf 440}, 69 (1998).
\bibitem{Raf96}
     G. G. Raffelt, {\it Stars as Laboratories for Fundamental Physics}
     (The University of Chicago Press, Chicago, 1996).
\bibitem{Eng90}
  J. Engel, D. Seckel, and A. C. Hayes, Phys. Rev. Lett. {\bf 65}, 960 (1990).
\bibitem{Raf99}
  See, e.g., G. G. Raffelt, Ann. Rev. Nucl. Part. Sci. {\bf 49}, 163 (1999).
\bibitem{Avi88}
     F. T. Avignone III {\it et al}., Phys. Rev. D {\bf 37}, 618 (1988);
     W. C. Haxton and K. Y. Lee, Phys. Rev. Lett. {\bf 66}, 2557 (1991).
\bibitem{Sre85}
     M. Srednicki, Nucl. Phys. {\bf B260}, 689 (1985).
\bibitem{Cha99}
     S. Chang, S. Tazawa, and M. Yamaguchi,
     Phys. Rev. D {\bf 61}, 084005 (2000). 
\bibitem{Han01}
     S. Hannestad and G. G. Raffelt, Phys. Rev. Lett. {\bf 88}, 071301
     (2002).
\bibitem{Dil00}
     L. Di Lella, A. Pilaftsis, G. Raffelt, and K. Zioutas,
     Phys. Rev. D {\bf 62}, 125011 (2000).
 \bibitem{Gui99}
     G. F. Guidice, R. Rattazzi, and J. D. Wells,
     Nucl. Phys. {\bf B544}, 3 (1999); 
     T. Han, J. D. Lykken, and R. J. Zhang,
     Phys. Rev. D {\bf 59}, 105006 (1999).      


\end{thebibliography}
\end{document}